\documentclass[proof]{WileyASNA-v1}

\articletype{Article Type}%

\received{xx xxxx xxxx}
\revised{xx xxxx xxxx}
\accepted{xx xxxx xxxx}

\begin{document}

\title{The gaseous natal environments
of GPS and CSS sources with ASKAP -- FLASH\protect}

\author[1,2]{James R. Allison}

\author[2,3,4]{Elaine M. Sadler}

\author[4]{Elizabeth K. Mahony}

\author[4,3]{Vanessa A. Moss}

\author[2,3]{Hyein Yoon}

\authormark{James R. Allison \textsc{et al}}

\address[1]{\orgdiv{Sub-Dept. of Astrophysics, Department of Physics}, \orgname{University of Oxford}, \orgaddress{\state{Denys Wilkinson Building, Keble Rd., Oxford, OX1 3RH}, \country{U.K.}}}

\address[2]{ARC Centre of Excellence for All-Sky Astrophysics in 3 Dimensions (ASTRO 3D)}

\address[3]{\orgdiv{Sydney Institute for Astronomy, School of Physics A28}, \orgname{University of Sydney}, \orgaddress{\state{NSW 2006}, \country{Australia}}}

\address[4]{\orgdiv{Space \& Astronomy}, \orgname{CSIRO}, \orgaddress{\state{PO Box 76, Epping, NSW 1710}, \country{Australia}}}

\corres{*James R. Allison,  \email{james.allison@physics.ox.ac.uk}}


\abstract{GPS and CSS sources are thought to represent a young and/or confined sub-population of radio-loud active galactic nuclei (AGN) that are yet to evacuate their surrounding ambient interstellar gas. By studying the gaseous environments of these objects we can gain an insight into the inter-dependent relationship between galaxies and their supermassive black holes (SMBHs). The First Large Absorption Survey in \mbox{H\,{\sc i}} (FLASH) will build a census of the neutral atomic hydrogen (\mbox{H\,{\sc i}}) gas in galaxies at intermediate cosmological redshifts. FLASH is expected to detect at least several hundred \mbox{H\,{\sc i}} absorbers associated with GPS and CSS sources. These absorbers provide an important probe of the abundance and kinematics of line-of-sight neutral gas towards radio AGN, in some cases revealing gas associated with infalling clouds and outflows. Observations are now complete for the first phase of the FLASH Pilot Survey and early analysis has already yielded several detections, including the GPS source PKS\,2311$-$477. Optical imaging of this galaxy reveals an interacting system that could have supplied the neutral gas seen in absorption and triggered the radio-loud AGN. FLASH will provide a statistically significant sample with which the prevalence of such gas-rich interactions amongst compact radio galaxies can be investigated.} 


\jnlcitation{\cname{%
\author{Allison J. R.}, 
\author{Sadler E. M.}, 
\author{Mahony E. K.}, and 
\author{Moss V.A.}} (\cyear{2021}), 
\ctitle{The gaseous natal environments of GPS and CSS sources with ASKAP-FLASH}, \cjournal{Astronomiche Nachrichten}, \cvol{xx;xx:x--x}.}


\maketitle


\section{Introduction}\label{section:introduction}

\vspace{-6pt}

The GHz-Peaked Spectrum (GPS) and Compact Steep Spectrum (CSS) radio sources are thought to be a young and/or confined sub-population, with linear sizes less than an about 20\,kpc (\citealt{O'Dea:2021}). As such, they provide excellent objects with which to study the interaction between radio-loud active galactic nuclei (AGN) and their host galaxies. In some cases multi-phase gaseous outflows are detected, possibly pointing to a form of radio-jet feedback in the early stages of radio AGN evolution (e.g. \citealt{Holt:2008}). The \mbox{H\,{\sc i}} 21-cm line, when detected as absorption in the radio spectrum, is a particularly useful method of studying the natal gaseous environments of young radio galaxies. Detection rates of \mbox{H\,{\sc i}} absorption amongst GPS and CSS sources tend to be relatively high ($\sim 30$\,per\,cent) and provide line-of-sight kinematic information on the cool neutral gas towards the source (see \citealt{Morganti:2018} for a review). However, the historical number of compact sources searched for \mbox{H\,{\sc i}} absorption remains insufficient to draw strong conclusions about the population, particularly at cosmological distances. Larger spectroscopic radio surveys with the new Square Kilometre Array (SKA) pathfinder telescopes are therefore required to build a statistically significant sample.

\section{The First Large Absorption Survey in \mbox{H\,{\sc i}} (FLASH)}\label{section:flash}

The First Large Absorption Survey in \mbox{H\,{\sc i}} (FLASH; \citealt{Allison:2021}) is a radio survey with the Australian Square Kilometre Array Pathfinder (ASKAP; \citealt{Hotan:2021}) to detect the 21-cm absorption-line towards radio sources south of $\delta \approx +40\deg$, covering \mbox{H\,{\sc i}} redshifts between $z = 0.4$ and $1.0$. ASKAP is a 36-dish interferometer that has an instantaneous bandwidth of 288\,MHz and a 31\,$\deg^{2}$ field of view at 800\,MHz. The FLASH sensitivity ($3 - 5$\,mJy\,beam$^{-1}$ per 18.5\,kHz channel) and total sky area ($\approx 34\,000\deg^{2}$) are such that we expect to detect more than a thousand \mbox{H\,{\sc i}} absorbers, including the host galaxies of radio-loud AGN. Of these associated \mbox{H\,{\sc i}} absorbers we expect that at least several hundred will be CSS and GPS radio sources, representing an order of magnitude increase over the current literature sample. 

\vspace{-12pt}

\subsection{FLASH Pilot Survey}\label{section:pilot survey}

The first phase of the FLASH Pilot Survey is now complete and data analysis is underway. Observations comprised mostly 2-hr pointings that reached FLASH sensitivity over a total sky area of approximately 1000\,$\deg^{2}$ (i.e. about 3\,per\,cent of the full survey). To enable rapid interpretation of detected \mbox{H\,{\sc i}} absorbers, the pilot fields were selected from existing surveys at optical wavelengths. These surveys include the Sloan Digital Sky Survey (SDSS) Baryon Oscillation Spectroscopic Survey (BOSS; \citealt{Dawson:2013}), the Galaxy And Mass Assembly survey (GAMA; \citealt{Liske:2015}), the Molonglo Reference Catalogue (\citealt{McCarthy:1996}) and the Dark Energy Survey (DES; \citealt{DES:2005}). Full details of the observations, data processing and results from the FLASH Pilot Survey will be given in a forthcoming paper by Yoon et al. (in preparation).

\vspace{-12pt}

\subsection{PKS\,2311$-$477}\label{section:examples}

\begin{figure}
\centerline{\includegraphics[width=0.5\textwidth]{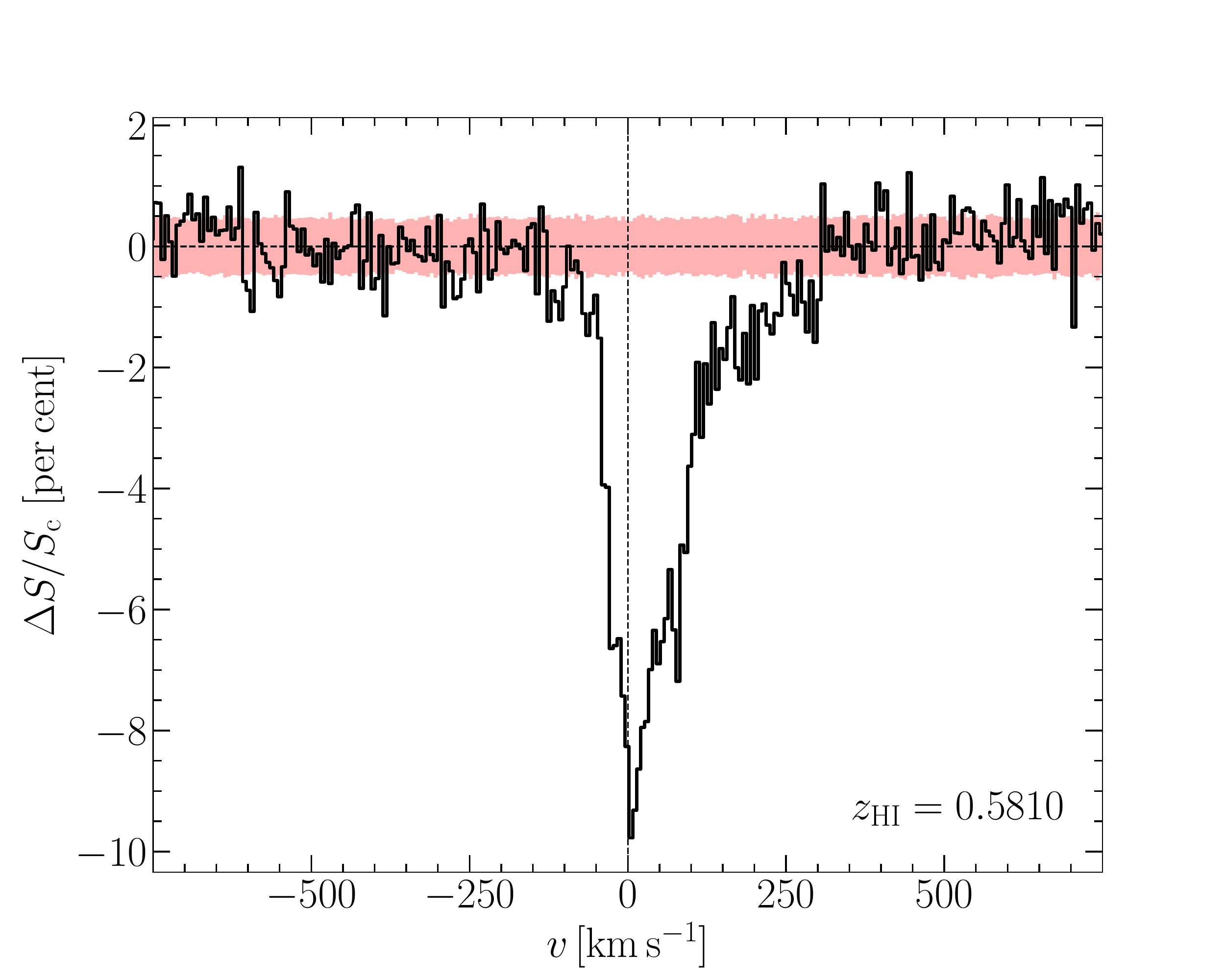}}
	\caption{An example \mbox{H\,{\sc i}} 21-cm absorption spectrum towards PKS\,2311$-$477, from ASKAP observations during the FLASH Pilot Survey. The flux density (black line) has been subtracted and normalised by the source continuum. The shaded red region denotes the measured rms noise per 18.5\,kHz channel. The velocity axis is given with reference to the peak absorption at a redshift of $z = 0.5810$.}\label{figure:PKS2311-477_ASKAP_spectrum}
\end{figure}

As an example, we present here a FLASH detection of \mbox{H\,{\sc i}} absorption towards the GPS radio source PKS\,2311$-$477. This source was observed with ASKAP during the FLASH Pilot Survey\footnote{Publicly released ASKAP data can be found at \url{https://research.csiro.au/casda/}. PKS\,2311$-$477 was observed in scheduling block SBID 15873.}. Details of the observations and data processing, which uses the ASKAPsoft pipeline (\citealt{Wieringa:2020}), will be discussed in detail by Yoon et al. (in preparation). A section of the ASKAP spectrum, centred on the detected \mbox{H\,{\sc i}} absorption line at $z = 0.5810$, is shown in \autoref{figure:PKS2311-477_ASKAP_spectrum}. The line has a peak optical depth of $\tau \approx 0.1$ and FWHM of $\Delta{v} \approx 150$\,km\,s$^{-1}$, with an asymmetric tail that extends to about +300\,km\,s$^{-1}$ from the peak. 

\begin{figure}
\centerline{\includegraphics[width=0.5\textwidth]{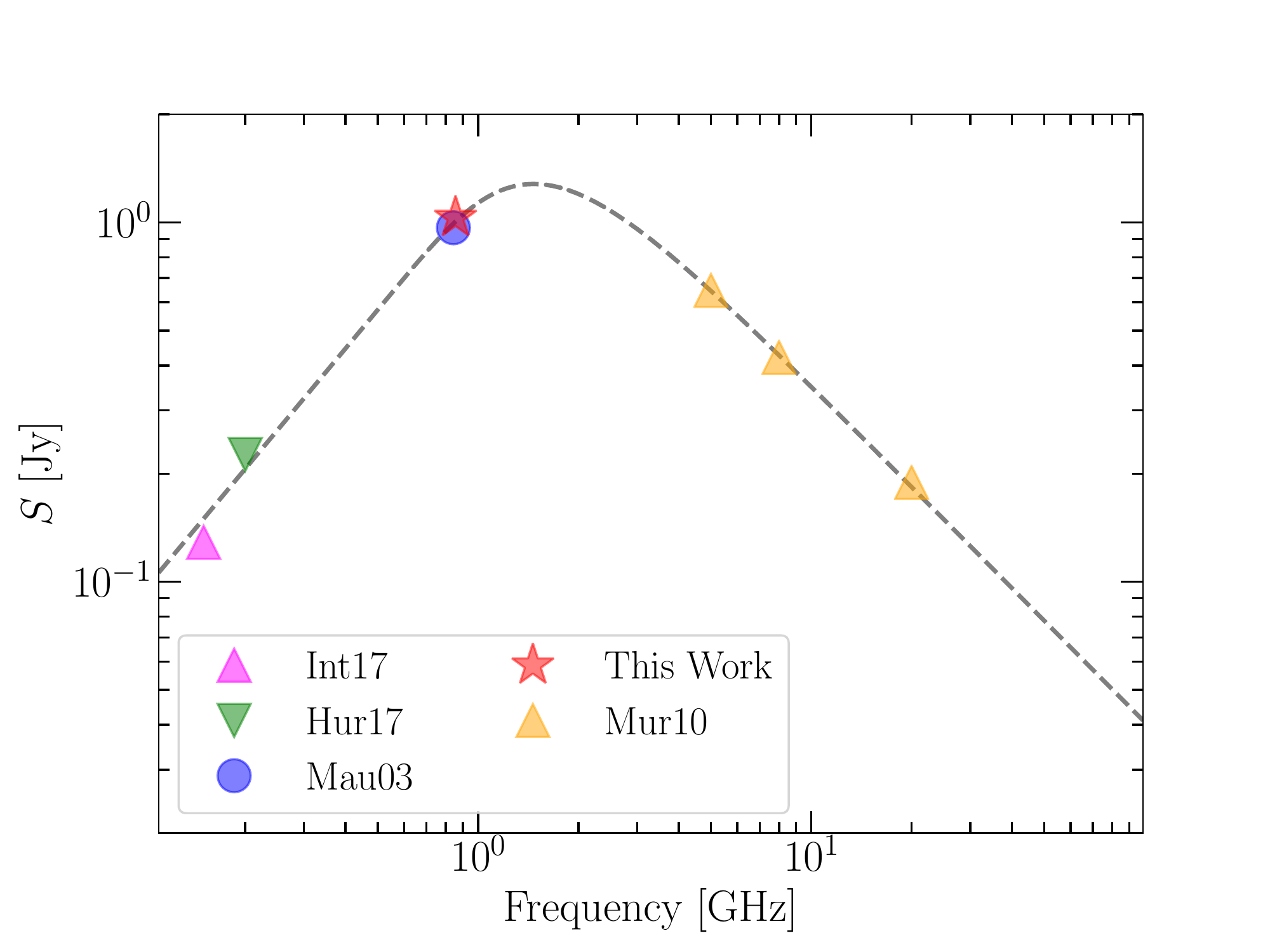}}
	\caption{The spectral energy distribution (SED) of PKS\,2311$-$477 at radio wavelengths, compiled using data from the literature and this work. The frequency axis is given in the observer rest frame. The dashed line denotes a best-fitting model that includes optically thick and thin power-law spectra (\citealt{Snellen:1998}). References for the data: Int17 -- \citet{Intema:2017}; Hur17 -- \citet{Hurley-Walker:2017};  Mau03 -- \citet{Mauch:2003}; Mur10 -- \citet{Murphy:2010}.}\label{figure:PKS2311-477_radio_sed}
\end{figure}

The spectral energy distribution (SED) of the background radio source is shown in \autoref{figure:PKS2311-477_radio_sed}. Using the least-squares method we fit the analytical GPS model of \cite{Snellen:1998}, obtaining a peak flux density of $S_{\rm peak} = 1.3$\,Jy at 1.6\,GHz and optically thick and thin spectral indices of $\alpha_{\rm tk} = 1.1$ and $\alpha_{\rm tn} = -0.9$, respectively. Based on this best-fitting SED model, we use the redshift of the \mbox{H\,{\sc i}} absorption-line to obtain a lower limit for the 1.4\,GHz radio luminosity of $L _{1.4} \gtrsim 1.0 \times 10^{27}$\,W\,Hz$^{-1}$, which is typical of powerful GPS sources.

As yet no optical spectroscopic information could be found for this source and so the absorber association remains unknown. However, the line width is more consistent with the width distribution for detected associated/intrinsic absorbers than intervening absorbers. We use recent machine learning analysis by \cite{Curran:2021} of the literature sample to estimate that there is about 80\,per\,cent probability that this absorber is associated with the host galaxy of PKS\,2311$-$477. The line profile is also consistent with that of \mbox{H\,{\sc i}} absorbers detected in luminous compact radio galaxies, which tend to be more asymmetric and broader, possibly tracing the irregular kinematics of neutral gas influenced by proximity to the radio source and AGN (\citealt{Gereb:2015}, \citealt{Maccagni:2017}). To verify this interpretation we are currently pursuing optical spectroscopy of this object as part of a larger follow up program of the FLASH Pilot Survey. 

\begin{figure}
\centerline{\includegraphics[width=0.5\textwidth]{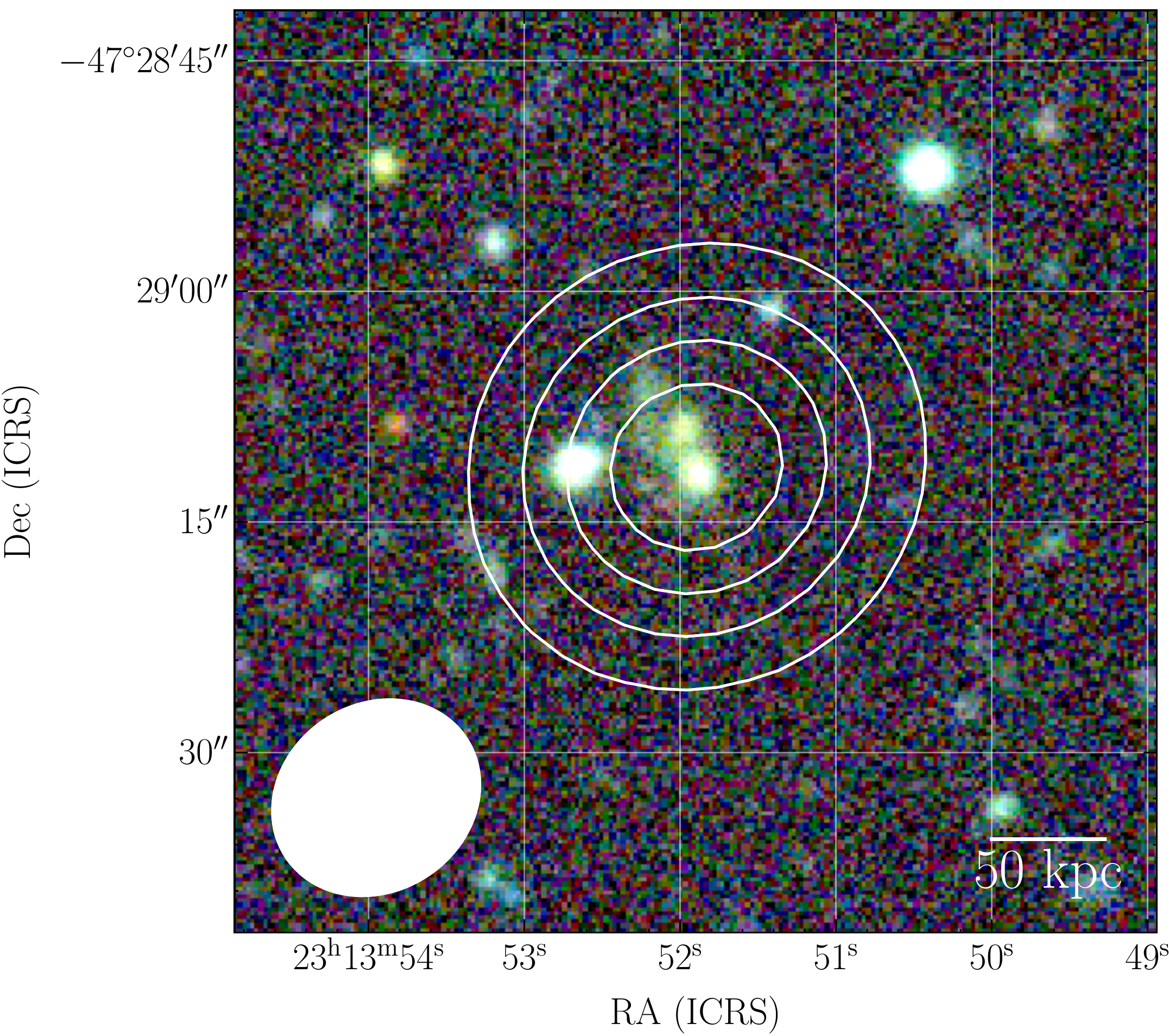}}
	\caption{An optical image of PKS\,2311$-$477, constructed using $grz$-band images obtained from the DESI Legacy Imaging Surveys (\citealt{Dey:2019}). The white contours show the 856\,MHz continuum from FLASH observations (levels are 0.1, 0.2, 0.3, and 0.4\,Jy\,beam$^{-1}$). The restoring beam is shown in the bottom left hand corner. The horizontal bar indicates the physical scale at the redshift of the \mbox{H\,{\sc i}} absorption line.}\label{figure:PKS2311-477_image}
\end{figure}

In \autoref{figure:PKS2311-477_image} we show a $grz$-band optical image of PKS\,2311$-$477, overlaid with contours showing the 856\,MHz continuum from FLASH. The optical image reveals at least two interacting galaxies at the position of the radio source; such interactions are thought to be progenitors of luminous radio galaxies and, if gas-rich, could be the reason why we detect a significant reservoir of \mbox{H\,{\sc i}} gas towards this source (e.g. \citealt{RamosAlmeida:2012, Chiaberge:2015}).

\vspace{-12pt}

\section{Summary}\label{section:summary}

FLASH is a wide-field survey for \mbox{H\,{\sc i}} 21-cm absorption using the ASKAP radio telescope, covering intermediate cosmological redshifts between $z = 0.4$ and $1.0$ (\citealt{Allison:2021}). The survey is expected to detect at least several hundred absorbers in the host galaxies of GPS and CSS sources, providing a statistically significant radio-selected sample with which to study the interaction between young and/or confined radio galaxies and their ambient environments. Spanning a broad range of intermediate cosmological redshifts, and by comparing with previous and contemporaneous \mbox{H\,{\sc i}} absorption surveys at other redshifts, this sample will be used to determine if there is evidence for late-time evolution in the population of these objects. 

Observations for the first 1000\,$\deg^{2}$ of the FLASH Pilot Survey are now complete (Yoon et al. in preparation) and data analysis is underway. Several absorption lines have been confirmed so far, including \mbox{H\,{\sc i}} absorption towards the GPS radio source PKS\,2311$-$477. Although no optical spectroscopic redshift yet exists for this source, the width and asymmetry of the line profile is consistent with neutral gas associated with the host galaxy of a luminous compact radio galaxy. Optical images available from the DESI Legacy Imaging Surveys (\citealt{Dey:2019}) show that PKS\,2311$-$477 is undergoing a significant merger or interaction with at least one other galaxy. We expect that the full FLASH survey will discover many more such systems, allowing us to test whether galaxy-galaxy interactions are an important feature in supplying the gas required to trigger such luminous young and/or confined radio galaxies.

\vspace{-12pt}


\section*{Acknowledgments}

JRA acknowledges support from a Christ Church Career Development Fellowship. Parts of this research were conducted by the Australian Research Council Centre of Excellence for All-sky Astrophysics in 3D (ASTRO 3D) through project number CE170100013.

The Australian SKA Pathfinder is part of the Australia Telescope National Facility which is managed by CSIRO. Operation of ASKAP is funded by the Australian Govern- ment with support from the National Collaborative Research Infrastructure Strategy. ASKAP uses the resources of the Pawsey Supercomputing Centre. Establishment of ASKAP, the Murchison Radio-astronomy Observatory and the Pawsey Supercomputing Centre are initiatives of the Australian Government, with support from the Government of Western Australia and the Science and Industry Endowment Fund. We acknowledge the Wajarri Yamatji people as the traditional owners of the Observatory site.

The Legacy Surveys consist of three individual and complementary projects: the Dark Energy Camera Legacy Survey (DECaLS; Proposal ID \#2014B-0404; PIs: David Schlegel and Arjun Dey), the Beijing-Arizona Sky Survey (BASS; NOAO Prop. ID \#2015A-0801; PIs: Zhou Xu and Xiaohui Fan), and the Mayall z-band Legacy Survey (MzLS; Prop. ID \#2016A-0453; PI: Arjun Dey). DECaLS, BASS and MzLS together include data obtained, respectively, at the Blanco telescope, Cerro Tololo Inter-American Observatory, NSF's NOIRLab; the Bok telescope, Steward Observatory, University of Arizona; and the Mayall telescope, Kitt Peak National Observatory, NOIRLab. The Legacy Surveys project is honored to be permitted to conduct astronomical research on Iolkam Du'ag (Kitt Peak), a mountain with particular significance to the Tohono O'odham Nation.

We have made use of Astropy, a community-developed core PYTHON package for astronomy (\citealt{Astropy:2013, Astropy:2018}); \textsc{Aplpy}, an open-source plotting package for Python (\citealt{Robitaille:2012}); NASA's Astrophysics Data System Bibliographic Services; and the VizieR catalogue access tool operated at CDS, Strasbourg, France.

\vspace{-12pt}










\bibliography{flash_gpscss_2021.bib}%

\vspace{-12pt}

\section*{Author Biography}

\vspace{-12pt}

\begin{biography}{\includegraphics[width=60pt,height=80pt]{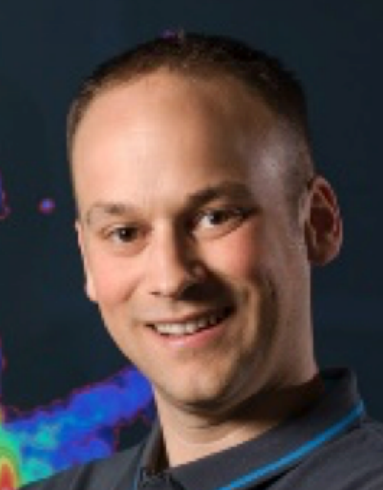}}{\textbf{James R. Allison.} The lead author is a Career Development Fellow in Astrophysics at Christ Church, the University of Oxford. He completed his DPhil in 2010, and since then has held research fellowships at Sydney, CSIRO and Oxford. He is currently a principal investigator of FLASH with the ASKAP telescope. In 2021 he participated in the 6th Workshop on Compact Steep Spectrum and GHz-Peaked Spectrum Radio Sources.}
\end{biography}

\end{document}